\documentclass[
 amsmath,
 aps,
 prl,
 twocolumn,
 superscriptaddress
]{revtex4-2}

\usepackage{xspace}
\usepackage[colorlinks]{hyperref}

\usepackage{gensymb}
\usepackage{float}
\usepackage{graphicx}
\usepackage{epstopdf}
\usepackage{setspace}
\usepackage{bbold}
\usepackage{comment}
\usepackage{color}
\usepackage{soul}
\usepackage{amsmath}
\usepackage{lipsum}
\usepackage[abs]{overpic}
\usepackage{nicefrac}
\usepackage{graphicx}
\usepackage{dcolumn}
\usepackage{bm}

\newcommand{\abs}[1]{\left\vert#1\right\vert}

\begin{document}

\title{Optomechanical Detection of Individual Gas Collisions}

\author{Yu-Han Tseng}
\thanks{Contact author: yuhan.tseng@yale.edu}
\affiliation{Wright Laboratory, Department of Physics, Yale University, New Haven, Connecticut 06520, USA}
\author{Clarke A. Hardy}
\thanks{Contact author: clarke.hardy@yale.edu}
\affiliation{Wright Laboratory, Department of Physics, Yale University, New Haven, Connecticut 06520, USA}
\author{T. W. Penny}
\altaffiliation{Present address: Department of Microtechnology and Nanoscience (MC2),
Chalmers University of Technology, 41296 Gothenburg, Sweden}
\affiliation{Wright Laboratory, Department of Physics, Yale University, New Haven, Connecticut 06520, USA}
\author{Cecily Lowe}
\affiliation{Wright Laboratory, Department of Physics, Yale University, New Haven, Connecticut 06520, USA}
\author{Jacqueline Baeza-Rubio}
\affiliation{Wright Laboratory, Department of Physics, Yale University, New Haven, Connecticut 06520, USA}
\author{Daniel Carney}
\affiliation{Physics Division, Lawrence Berkeley National Laboratory, Berkeley, California 94720, USA}
\author{David C. Moore}
\affiliation{Wright Laboratory, Department of Physics, Yale University, New Haven, Connecticut 06520, USA}
\affiliation{Yale Quantum Institute, Yale University, New Haven, Connecticut 06520, USA}
\date{\today}

\begin{abstract}
We experimentally demonstrate the detection of momentum transfers from individual collisions of Kr, Xe, and SF$_6$ with an optically levitated nanoparticle, finding good agreement with theoretical expectations.
The observed event rates accurately measure the gas partial pressures, while the spectral shape provides a sensitive probe of the surface properties of the nanoparticle, including its temperature.
The reconstruction of impulse signals as small as 200~$\mathrm{keV}/c$ further establishes that levitated optomechanical sensors can reach the sensitivity required for precision measurements of fundamental particle interactions, and demonstrates a proof-of-principle for a primary pressure sensor based on the detection of individual gas particle collisions.
\end{abstract}
\maketitle

\paragraph*{Introduction.}
A mechanical system interacting with its thermal environment is governed by forces that can be precisely described without detailed knowledge of their microscopic origin.
Fundamentally, however, these forces arise from interactions with microscopic degrees of freedom, which are typically too numerous and weak to resolve individually.
In Brownian motion, for example, the macroscopic behavior of a particle emerges from many unresolved collisions with atoms or molecules in a surrounding fluid~\cite{Brown1828, Einstein1905, Smoluchowski1906, li_measurement_2010}.
Here, we demonstrate that advances in levitated optomechanics can now enable the resolution of individual collisions between gas molecules and a nanoscale particle.

Detection of individual gas collisions with a macroscopic mechanical system is directly relevant to ongoing experimental programs studying such systems in the quantum regime~\cite{aspelmeyer_cavity_2014}.
While many sources of decoherence can be mitigated by levitation~\cite{gonzalez-ballestero_levitodynamics_2021}, uncontrolled information loss due to scattering from background gas molecules remains a fundamental challenge~\cite{joos_emergence_1985, hornberger_collisional_2003}.
Despite considerable progress in preparing and manipulating the quantum states of levitated particles~\cite{delic_cooling_2020, magrini_real-time_2021, tebbenjohanns_quantum_2021, rossi_quantum_2024, kamba_quantum_2025, dania_high-purity_2024, troyer_quantum_2026}, gas-induced decoherence remains a significant constraint for experiments that aim to study quantum mechanics at macroscopic scales~\cite{romero-isart_large_2011} and the quantum nature of gravity~\cite{2013arXiv1311.4558K, bose_spin_2017, marletto_gravitationally_2017}.
In addition, for proposed searches for new fundamental particles that utilize levitated sensors, residual gas collisions can introduce background events~\cite{afek_coherent_2022, carney_searches_2023}.
This work provides quantitative measurements that could inform the modeling of gas-induced backgrounds in these proposed experiments.

In this work we demonstrate the detection of impulsive forces in the 200--600~$\mathrm{keV}/c$ range arising from individual gas molecules scattering from an optically levitated nanoparticle.
Detection of these forces is enabled by recently demonstrated techniques for quantum noise-limited impulse measurement in such systems~\cite{tseng_search_2025, skrabulis_nanomechanical_2026}.
By injecting trace amounts of heavy gases, we quantitatively study the collisions between gas molecules and the nanoparticle on an event-by-event basis.
We show that the partial pressures of various gas species can be determined from the scattering rates.
Furthermore, the spectral shape provides a measurement of the surface properties of the trapped particle, including its temperature.
This measurement demonstrates the potential of levitated sensors for proposed applications to high-sensitivity pressure sensors~\cite{barker_collision-resolved_2024, gajewski_levitas_2025} and impulse-based searches for new physics, including dark matter~\cite{PhysRevLett.125.181102, afek_coherent_2022, Kilian:2024fsg, tseng_search_2025} and sterile neutrinos~\cite{carney_searches_2023}.

\begin{figure}
    \centering
    \includegraphics[width=\columnwidth]{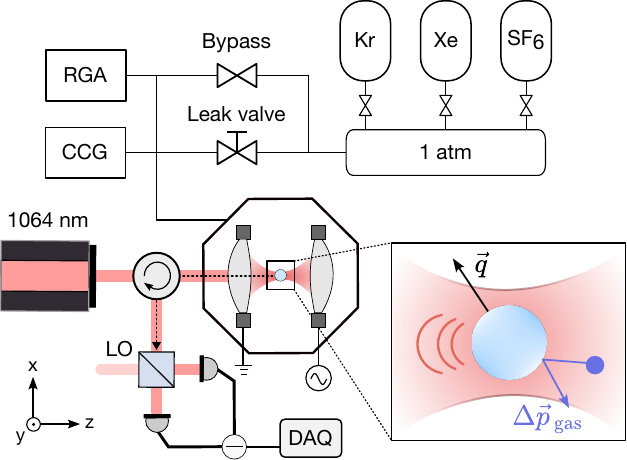}
    \caption{Simplified schematic illustrating the experimental setup. A silica nanoparticle is optically trapped in vacuum, and the $z$ position of the nanoparticle is continuously monitored. A gas handling system enabling controlled injection of Kr, Xe, and SF$_6$ is attached to the vacuum chamber, with a residual gas analyzer (RGA) and a cold cathode gauge (CCG) used to monitor the gas composition and pressure. The inset illustrates a gas particle scattering from the trapped nanosphere and imparting an instantaneous momentum transfer, $\vec{q}$. 
    }
    \label{fig:gas_drawing}
\end{figure}

\paragraph*{Experimental description.}
A silica nanosphere (nominal radius $R=50$~nm with mass $m \approx 1~\mathrm{fg}$) is optically trapped in ultrahigh vacuum using a 1064~nm laser focused by an aspheric lens (numerical aperture 0.77).
See Fig.~\ref{fig:gas_drawing} and Ref.~\cite{tseng_search_2025} for a detailed description of the experimental configuration.
The optical trap forms a three-dimensional harmonic potential for the center-of-mass motion of the nanoparticle, which is cooled through feedback damping by parametric modulation of the laser power~\cite{PhysRevLett.109.103603}.
Throughout the measurement described in this work, the nanoparticle is trapped at a background pressure of $(3 \pm 1) \times 10^{-8}$~mbar where H$_2$O and H$_2$ are the dominant residual gases measured with a residual gas analyzer (RGA).
The net electric charge of the particle is monitored using an off-resonance electric field oscillating at 137~kHz and is maintained at +$6~e$ throughout data collection for the \textit{in situ} impulse calibration described below.
The off-resonance electric signal is also used to monitor the stability of the homodyne detection over time.

Here, we focus on the nanoparticle's motion in $z$ (aligned with the direction of the laser propagation), which is well described by a damped harmonic oscillator with a resonant frequency $\Omega_z / 2 \pi = 48 \pm 1~\mathrm{kHz}$.
The phase of the light that is backscattered from the nanoparticle is interferometrically read out using balanced homodyne detection and digitized at a sampling rate of 5~MS/s, providing a sensitive measurement of the $z$ position~\cite{tebbenjohanns_optimal_2019, maurer_quantum_2022}.

A gas handling system that enables controlled injection of Kr, Xe, and SF$_6$ [atomic or molecular masses 83.798(2), 131.293(6), and 146.055(5)~u] is attached to the vacuum chamber.
During data taking, gas is injected by pressurizing the gas handling manifold to $\approx$1~atm and opening the leak valve until a cold cathode gauge (CCG) attached to the manifold reads the desired pressure.
The purities of the injected gases are estimated to be $>$99.99\%.
A turbomolecular pump and a nonevaporable getter are connected directly to the vacuum chamber, maintaining a steady-state pressure set by the leak rate throughout each measurement.

On the microscopic level, an incoming gas particle interacting with the surface of the trapped nanosphere imparts a momentum, $\vec{q}$, over the residence time, $\tau_{\mathrm{res}}$, before leaving the surface.
The exact dynamics depends on the surface properties and the gas under consideration~\cite{Goodman1976}.
For all of the gas species considered in this work, $\tau_\mathrm{res} \ll 2 \pi / \Omega_z \approx 20~\mu \mathrm{s}$ and the collision events are well approximated by impulsive, instantaneous momentum transfers.
The response of the nanoparticle is calibrated by applying electric impulses in the $z$ direction with amplitudes spanning $100$--$1000~\mathrm{keV}/c$ via electrodes surrounding the optical trap, following the procedure established for similar systems~\cite{wang_mechanical_2024, tseng_search_2025}.
A matched filter is used to reconstruct the impulse amplitudes from the $z$ position measurement, where the signal is modeled as the response of a driven damped harmonic oscillator to a broadband force~\cite{tseng_search_2025}.
The reconstructed amplitudes are found to be linearly related to the known amplitudes of the applied impulses across the entire calibration range, with resolutions $\sigma_q=40$--$60~\mathrm{keV}/c$ measured for pulses well above the noise level.
The measured resolutions are within a factor of 4--6 of the standard quantum limit for impulse measurement~\cite{PhysRevB.70.245306}, $\Delta p = \sqrt{\hbar m \Omega_z} \approx 11~\mathrm{keV}/c$, for this system.

For each of the three gas species, datasets were collected at seven pressures ranging from $(5 \pm 2)\times10^{-8}$ to $(1.4 \pm 0.5)\times10^{-6}$~mbar, as indicated by the CCG readings.
The data for each pressure consist of a continuous measurement of the $z$ position of the nanoparticle with a live time of 2.8 min.
Before and after the injection of each gas species, a background dataset is collected with the leak valve fully closed and the vacuum chamber pumped to its base pressure.
Dedicated impulse calibrations spanning the full analysis range are performed before and after each gas injection, and regular $1.04 \pm 0.05 ~\mathrm{MeV}/c$ electric impulses that provide \textit{in situ} monitoring of the detector response are applied every 300~ms during data collection.

\begin{figure}
    \centering
    \includegraphics[width=\columnwidth]{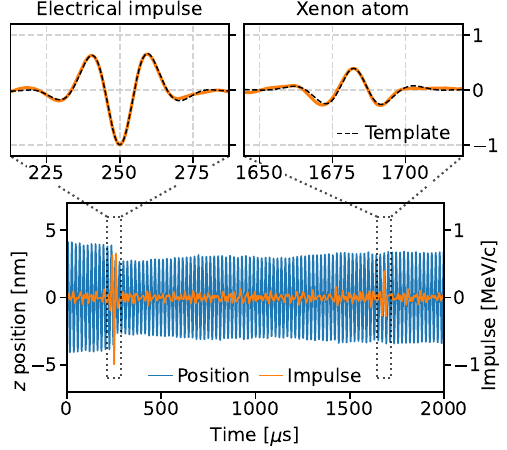}
    \caption{Lower plot: the $z$ position of the nanosphere (blue) and the reconstructed impulse amplitude (orange) shown over a 2~ms window. Two impulse events $>200~\mathrm{keV}/c$ are visible and are enlarged in the callouts. Left callout: reconstructed waveform of a nominally $1.04 \pm 0.05~\mathrm{MeV}/c$ electric calibration impulse. Right callout: a candidate xenon collision event with a measured amplitude of $390 \pm 60~\mathrm{keV}/c$. The signal template, derived by averaging the reconstruction of 800 calibration impulses, is shown in both callouts as a dashed black line.
    }
    \label{fig:sample_impulses}
\end{figure}

\paragraph*{Impulse measurement.}
For each dataset, the impulse amplitudes are reconstructed from the position measurement following the same procedure applied to the calibration data.
Matched filters are constructed separately in 100~ms-long analysis windows to account for slow drift in the oscillator's frequency, and the maximum reconstructed amplitude in each 50~$\mu$s-long search window is recorded.
This trigger-free search strategy accurately reconstructs the amplitudes of randomly arriving impulses that are well above the noise level.
The reconstructed waveforms of example impulse events are shown in Fig.~\ref{fig:sample_impulses}.

Data selection cuts are applied to reject data that show evidence of unmodeled transient noise or significant drift in the detector response~\cite{tseng_search_2025}.
A noise-level cut rejects noisy time windows for which the rms reconstruction amplitude in the surrounding 250~$\mu$s (excluding the detected impulse) exceeds the mean level by $>1\sigma$.
A detection stability cut is then applied to reject search windows in which the measured power of the off-resonance monitoring signal falls more than $1\sigma$ below the mean.
On average, these cuts remove $30\%$ of the live time, but minimize systematic errors in the reconstructed spectral shape arising from time-dependent drifts.
Finally, an event-level selection is applied based on the goodness of fit of each reconstructed waveform to the template derived from calibration data, further removing 2\% of impulse candidates.
Monte Carlo simulations and calibration data are used to verify that these data selection cuts do not distort the signal spectra or introduce biases in the parameter estimation presented in the next section.

All of the $1.04~\mathrm{MeV}/c$ calibration impulses applied during data taking are found to be correctly reconstructed and are excluded from the subsequent analysis based on timing information. 
Candidate impulse events that pass all selection criteria are binned into histograms, which are then converted to differential event rates after correcting for the effective live time.

\begin{figure*}
\includegraphics[width=\textwidth]{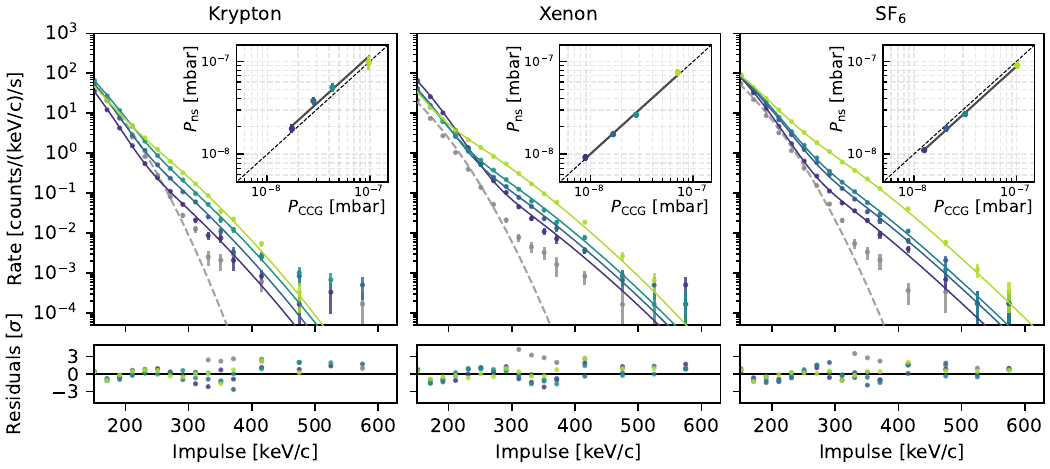}
    \caption{Measured impulse event rates for Kr (left), Xe (center), and SF$_6$ (right). The colored points (in order of pressure increasing from darkest to lightest) show the measured rates with gas present, while the solid lines are fits to the sum of the signal and background models. Background-only data (no gas injected) are shown in gray with Gaussian fits shown as dashed lines. Normalized residuals are shown in the panel below each plot. Each of the datasets contains $\approx 200000$ events above the analysis threshold, representing 2.8~min of raw data before data selection cuts. The insets show the partial pressures of the injected gases fitted from the nanosphere data, $P_\mathrm{ns}$, compared to the zero-corrected cold cathode gauge readings, $P_\mathrm{CCG}$, where the subtracted base pressure is extracted from a linear fit. The solid and dashed lines indicate linear fits to the data and the ideal case where $P_\mathrm{ns} = P_\mathrm{CCG}$, respectively. All error bars are uncertainties calculated from the likelihood model and do not include the shared systematic errors described in the text.}
    \label{fig:combined_gas_spectra} 
\end{figure*}

Figure~\ref{fig:combined_gas_spectra} shows the measured impulse event rates for the three injected gas species at various pressures.
In all datasets, the measured events with amplitudes ${\lesssim 200~\mathrm{keV}/c}$ are dominated by the Gaussian reconstruction noise (with a width $\sigma \approx 60~\mathrm{keV}/c$, varying slightly between datasets).
Reconstruction for amplitudes $\lesssim 100~\mathrm{keV}/c$ is biased by searching for the maximum signal in the presence of noise~\cite{tseng_search_2025}.
For each gas species, an event population that grows with the increasing injected partial pressure is observed above $\approx200~\mathrm{keV}/c$.
These impulses, with rates falling off rapidly with increasing impulse amplitude, represent the candidate gas collision events.
Excess events that show up as non-Gaussian tails with a structure similar to the candidate gas-collision distributions, but at much lower rates, are observed in the background datasets.
Although future work is required to identify the origin of these background events, they could arise from collisions with residual background gas or time-dependent environmental noise that is not fully mitigated~\cite{tseng_search_2025}.

\paragraph*{Data analysis.}
We use a binned likelihood model to perform a joint fit over datasets for each gas to show that the measured event rate spectra agree quantitatively with a simple model based on the kinetic theory of gases.
The number of counts in each impulse bin is described by a Poisson process, though in our data we observe $\approx5\%$ excess variance attributable to data quality cuts and fluctuations in the detector response.
To model this, we use a negative binomial likelihood~\cite{lawless_negative_1987,breslow_tests_1990} with 5\% overdispersion.
Each bin includes contributions from the Gaussian background and a theoretical signal model, with the sum over all bins constrained to match the total number of events.
An analysis threshold of $150~\mathrm{keV}/c$ is used, which is well above the reconstruction bias due to the search effect, as described in the previous section.

The signal model is calculated following Ref.~\cite{barker_collision-resolved_2024} by integrating the velocity distributions of incoming and outgoing gas particles over the surface area of the nanoparticle, including both specular (elastic) and diffuse scattering.
In the latter case, the gas particles thermalize with the surface of the nanosphere before being emitted in a random direction.
The thermal accommodation coefficient, $\alpha$, which is a property of the gas-surface interaction dynamics, sets the fraction of scatters that are diffuse.
The differential scattering rate along the $z$ axis can be written as a weighted sum of the specular scattering rate, $d\Gamma_s / d|q_z|$, and the diffuse scattering rate, $d\Gamma_d / d|q_z|$, 
\begin{equation}
    \label{eq:rate}
    \frac{d\Gamma}{d|q_z|} (R, P, m_g, \alpha, T_g, T_s) = 
    \left(1-\alpha\right) \frac{d\Gamma_s}{d|q_z|} + \alpha\frac{d\Gamma_d}{d|q_z|}.
\end{equation}
Here $q_z$ is the $z$ component of the momentum transfer. The specular scattering rate is a function of the nanosphere radius ($R$), the gas pressure ($P$), the mass of a gas particle ($m_g$), and the gas temperature ($T_g$; $T_g=293~\mathrm{K}$ is used throughout this work).
In addition to the aforementioned parameters, the diffuse scattering rate further depends on the sphere surface temperature ($T_s$).
Detailed mathematical forms of $d\Gamma_s / d|q_z|$ and $d\Gamma_d / d|q_z|$ are provided in End Matter.
The effect of the measurement resolution is incorporated by convolving the event rates calculated from Eq.~\eqref{eq:rate} with a Gaussian kernel of width $\sigma_q$.

In the joint fit for each gas species, the pressure and detector resolution are free to vary between datasets taken at different pressures, while $\alpha$ and $T_s$ are common to all.
The detector resolution for each dataset is constrained in the fit by a Gaussian term in the likelihood centered at the resolution estimated from the calibration impulses applied during data taking, and with a 1$\sigma$ width (as a fraction of the mean) of 10\%.
Data taken at partial pressures greater than $10^{-7}$~mbar are excluded from the analysis, as non-negligible distortions of signal spectra due to event pile up are expected at these pressures based on simulation studies.
To estimate the partial pressure, the parameters $\sigma_q$, $T_s$, and $\alpha$ are treated as nuisance parameters and are marginalized over in the fit.
The best-fit pressures are then compared to the pressures reported by the CCG, with an appropriate gas-dependent correction factor applied to the raw CCG readings.

\paragraph*{Results and discussion.}
We find the best-fit partial pressures from the observed collision spectra agree with those reported by the CCG in the $10^{-8}$--$10^{-7}$~mbar range, to within uncertainties (Fig.~\ref{fig:combined_gas_spectra} insets).
Fitting the same signal models to the background-only datasets indicates a minimum resolvable partial pressure of $\approx2\times10^{-9}$~mbar.
To probe partial pressures beyond the current limit, accurate modeling or mitigation of events in the non-Gaussian tail of the background-only spectra is required.
An improved momentum resolution, which can be achieved by further optimization of the position readout or manipulation of the nanoparticle's motional state~\cite{skrabulis_nanomechanical_2026, Marocco:2026gwa, otabe2026timeofflightforcesensingquantum}, can substantially reduce the contribution of the Gaussian reconstruction noise at smaller impulse amplitudes, extending the measurement technique to lower pressures and lighter gas species.
The absolute scale of our pressure estimates is dominated by systematic uncertainties in the electric field configuration ($\lesssim5\%$ as estimated from finite element analysis) and in the sphere surface area ($\lesssim 10\%$ as reported by the manufacturer).
The CCG readings are also subject to a common systematic uncertainty of $\approx30\%$.

We also extract the sphere surface temperature and the thermal accommodation coefficient from the fits, as these parameters affect the spectral shape in distinguishable ways.
Monte Carlo simulations with synthetic impulse waveforms injected into realistic noise data are used to verify that the analysis procedure used here can accurately reconstruct both parameters. 

At the pressures considered in this work, the heat transfer to gas particles through collisions is negligible.
Consequently, the surface temperature of the nanoparticle is set by the equilibrium between the blackbody exchange with the room temperature chamber walls and absorption of the trapping laser.
This sets a physical lower bound on $T_s$ at room temperature, whereas $\alpha$ is bounded between 0 and 1 by definition.
We therefore impose minimal priors of $T_s\geq293$~K and $0\leq\alpha\leq1$ and marginalize over the remaining parameters to obtain conservative Bayesian estimates in the presence of the boundaries.
For all three gases, the best-fit sphere surface temperature is found to lie at the lower bound, indicating negligible absorption of laser power.
We therefore report upper limits on $T_s$ in Table~\ref{tab:fit_results}, along with the measured values of $\alpha$, for each of the three gases.

Our results show broad consistency with existing measurements of the thermal accommodation coefficients for gases interacting with silica~\cite{bayer-buhr_determination_2022, daun_investigation_2008, yamaguchi_measurement_2014, ganta_optical_2011, sonnick_thermal_2020}.
A surface temperature of the nanoparticle close to 293~K, as suggested by the analysis described here, has potential implications for thermal decoherence in levitated systems~\cite{Hackermueller:2004ygx, romero-isart_large_2011}.
Following the approach described in Ref.~\cite{chang_cavity_2010} and using refractive index data from Ref.~\cite{kitamura_optical_2007}, we find that an equilibrium at room temperature requires an absorption coefficient of $\lesssim10~\mathrm{dB/km}$ at 1064~nm, consistent with existing measurements for amorphous silica \cite{loriette_absorption_2003, li_internal_2025}.
However, higher absorptions that cause elevated surface temperatures have also been reported for levitated silica particles~\cite{monteiro_optical_2017, millen_nanoscale_2014, hebestreit_measuring_2018, zhang_determining_2023}.
Future studies in which the nanoparticle and the injected gas are selectively heated during the measurement can further improve the surface temperature measurement and quantify possible sphere-to-sphere variations.

\begin{table}
\caption{\label{tab:fit_results}Thermal accommodation coefficients ($\alpha$) and sphere surface temperatures ($T_s$) measured from joint fits to the impulse spectra for each gas. Uncertainties on $\alpha$ and $T_s$ are 68\% intervals and 95\% upper limits, respectively, computed with Markov Chain Monte Carlo sampling of the posterior probability distributions.}
\renewcommand{\arraystretch}{1.5}
\begin{ruledtabular}
\begin{tabular}{lcc}
Gas type &    $\alpha$ & $T_s$ {[}K{]}  \\ \cline{1-3}
Kr 	&    $0.55^{+0.11}_{-0.14}$    &    $<351$    \\
Xe 	&    $0.61^{+0.11}_{-0.12}$    &    $<353$    \\
SF$_6$ 	&    $0.82^{+0.07}_{-0.08}$    &    $<334$    \\
\end{tabular}
\end{ruledtabular}
\end{table}

\paragraph*{Conclusions.}
We have shown that impulsive forces from individual gas particles scattering from a silica nanosphere can be directly measured using techniques developed in levitated optomechanics.
With controlled injection of three different gas species, momentum transfers from the collisions are individually resolved beyond the continuous Brownian regime, with the amplitudes reconstructed on an event-by-event basis.
The measured event rate spectra are found to be in good agreement with theoretical predictions, providing accurate measurements of gas properties.

The detection of recoils from individual gas particles---which are approximately $10^{-7}$ times lighter in mass than the levitated sensor used here---demonstrates the potential of macroscopic mechanical systems in precision force sensing and the measurement of fundamental particle interactions~\cite{carney_mechanical_2021, carney_searches_2023}.
The direct observation of gas collisions can also inform models of collisional decoherence in experiments that target the creation of macroscopic quantum superposition in levitated systems.
Finally, the techniques demonstrated here may find applications in metrology, including the possibility of establishing a new pressure standard in the extreme-high vacuum regime~\cite{barker_collision-resolved_2024}.
For potential applications in pressure sensing, hollow particles with a large surface-to-mass ratio may be advantageous by simultaneously offering large event rates and a high momentum sensitivity.

\paragraph*{Acknowledgments.}
We thank Lorenzo Magrini, Daniel Barker, and the QuIPS team at Lawrence Berkeley National Lab for helpful discussions.
This work was supported by the U.S. DOE Office of Science, High Energy Physics, through the QuantISED 2.0 program, award DE-SC0026367, and in part by ONR Grant N00014-23-1-2600 and NSF Grant PHY-2512192. This work has been supported in part by the ``Table-top experiments for fundamental physics" program, sponsored by the Gordon and Betty Moore Foundation (Grant GBMF12328, DOI 10.37807/GBMF12328), Simons Foundation, Alfred P. Sloan Foundation (Grant G-2023-21130), and the John Templeton Foundation.
Y.-H.~T. is supported by the Graduate Instrumentation Research Award (GIRA) from the Coordinating Panel for Advanced Detectors (CPAD). J.~B.-R. and C.~L. are supported by the National Science Foundation Graduate Research Fellowship Program under Grant DGE-2139841.
D.~C. is supported by the U.S. DOE, Office of High Energy Physics, under Contract No. DEAC02-05CH11231, a DOE Early Career Research Award DE-SCL0000025, and by the Gordon and Betty Moore Foundation's Mid-Career Experimental Investigator Award GBMF13781.

\bibliography{references_gas}

\appendix

\section*{End Matter}

\paragraph*{Impulse reconstruction.}
The impulse signals, including those from gas collisions and electric calibration pulses, were reconstructed using the matched filtering technique described here.
With the position measurement, a matched filter is constructed to optimally weight the signal-to-noise power and provide a linear estimate of the amplitude of a known signal at each time.
For a recorded position waveform $z_n$, where $n = 0, 1, ..., N-1$ is the sample index in time, the reconstructed impulse amplitude is given by~\cite{tseng_search_2025}
\begin{equation}
A_n = \sum^{N-1}_{k=0} \frac{\tilde{s}^*_k \tilde{z}_k e^{2 \pi i k n / N}}{J_k} \propto \sum^{N-1}_{k=0} \frac{\chi^*_k \tilde{z}_k e^{2 \pi i k n / N}}{\abs{\chi_k}^2 + C_\mathrm{imp}}.
\label{eq:optimal_filter}
\end{equation}
Here $\tilde{s}_k$ and $\tilde{z}_k$ are the discrete Fourier transforms of the normalized signal model and $z_n$, respectively, and $J_k$ is the power spectral density describing the noise in the position measurement.
In Eq.~\eqref{eq:optimal_filter}, we have used $\tilde{s} \propto \chi$ for a broadband impulsive force and $J \propto (\abs{\chi}^2 + C_\mathrm{imp})$, where $\chi$ is the susceptibility of the oscillator and $C_\mathrm{imp}$ is the normalized imprecision noise floor.
The reconstructed amplitudes, $A_n$, are then calibrated into physical units using the known amplitudes of the calibration pulses.
We note that the reconstruction method applied here is acausal and is optimal when the noise is Gaussian and stationary, as opposed to techniques based on real-time state estimation~\cite{magrini_real-time_2021, schmerling_optimal_2024}.

\paragraph*{Signal model.}
Here, we provide the complete forms of the specular and diffuse scattering rates in Eq.~\eqref{eq:rate}.
We refer the readers to Ref.~\cite{barker_collision-resolved_2024} for the derivation of the formulas used here.
With the notations defined in the text, the specular scattering rate is given by
\begin{equation}
    \frac{d\Gamma_s}{d|q_z|} = \frac{\pi R^2 P}{m_gk_BT_g}\,\mathrm{erfc}\!\left(\!\frac{|q_z|}{\sqrt{8m_gk_BT_g}}\!\right),
\end{equation}
where $k_B$ is the Boltzmann constant. The diffuse scattering rate is
\begin{widetext}
\begin{equation}
\begin{aligned}
    \frac{d\Gamma_d}{d|q_z|}
    &= \frac{4\pi R^2 P}{m_g k_B T_g}\left(\frac{1}{2\pi\bar v^2}\right)^{1/2}
    \frac{\bar v^2}{\left(\bar v^2+\bar v_s^2\right)^2}\sqrt{\frac{\pi}{2}}
    \Bigg(\left[\bar v^{3}\operatorname{erfc}\!\left(\frac{|q_z|}{\sqrt{2}\,m_g\bar v}\right)
    +\bar v_s^{3}\operatorname{erfc}\!\left(\frac{|q_z|}{\sqrt{2}\,m_g\bar v_s}\right)\right] \\
    &\quad
    +\bar v\bar v_s\Bigg\{\bar v\operatorname{erfc}\!\left(\frac{|q_z|}{\sqrt{2}\,m_g\bar v_s}\right)
    +\bar v_s\operatorname{erfc}\!\left(\frac{|q_z|}{\sqrt{2}\,m_g\bar v}\right) \\
    &\quad
    +\sqrt{\bar v^2+\bar v_s^2}\,
    e^{-|q_z|^2/2m_g^2\left(\bar v^2+\bar v_s^2\right)}
    \left[\operatorname{erf}\!\left(\frac{|q_z|\,\bar v_s}{m_g\bar v\sqrt{2\left(\bar v^2+\bar v_s^2\right)}}\right)
    +\operatorname{erf}\!\left(\frac{|q_z|\,\bar v}{m_g\bar v_s\sqrt{2\left(\bar v^2+\bar v_s^2\right)}}\right)\right]\Bigg\} \\
    &\quad
    -\int_0^{\pi/2}\tan\theta\,d\theta\,\bar v\,\bar v_s\sqrt{\bar v^2+\bar v_s^2}\,
    \left[\operatorname{erf}\!\left(\frac{|q_z|\sec\theta\,\bar v_s}{m_g\bar v\sqrt{2\left(\bar v^2+\bar v_s^2\right)}}\right)
    +\operatorname{erf}\!\left(\frac{|q_z|\sec\theta\,\bar v}{m_g\bar v_s\sqrt{2\left(\bar v^2+\bar v_s^2\right)}}\right)\right]
    e^{-|q_z|^2\sec^2\theta/2m_g^2\left(\bar v^2+\bar v_s^2\right)}\Bigg),
\end{aligned}
\end{equation}
\end{widetext}
where $\bar v = \sqrt{k_B T_g / m_g}$ and $\bar v_s = \sqrt{k_B T_s / m_g}$ are the rms thermal velocity at $T_g$ and $T_s$, respectively.

\end{document}